\documentclass[twocolumn,showpacs,preprintnumbers,amsmath,amssymb]{revtex4}
\usepackage{graphicx}
\usepackage{dcolumn}
\usepackage{bm}
\usepackage{longtable}
\def\bea {\begin{eqnarray}}
\def\eea {\end{eqnarray}}

\def\be {\begin{equation}}
\def\ee {\end{equation}}

\begin{document}
\title{Crossing of large multiquasiparticle magnetic-rotation bands in $^{198}$Bi }
\author{H. Pai}
\altaffiliation{Present Address: Institut f\"ur Kernphysik, Technische Universit\"at Darmstadt, GERMANY}
\email{hari.vecc@gmail.com, hpai@ikp.tu-darmstadt.de, H.Pai@gsi.de}
\author{G. Mukherjee}
\thanks{Corresponding author}
\email{gopal@vecc.gov.in}
\author{S. Bhattacharyya}
\author{C. Bhattacharya}
\author{S. Bhattacharya}
\author{T. Bhattacharjee}
\author{S. K. Basu}
\author{S. Kundu}
\author{T. K. Ghosh}
\author{K. Banerjee}
\author{T. K. Rana}
\author{J. K. Meena}
\affiliation{Variable Energy Cyclotron Centre, Kolkata 700064, India}
\author{R. K. Bhowmik, R.P. Singh, S. Muralithar}
\affiliation{Inter University Accelerator Centre, Aruna Asaf Ali Marg, New Delhi 110067, INDIA}
\author{S. Chanda}
\affiliation{Fakir Chand College, Diamond Harbour, West Bengal, India }
\author{R. Garg}
\affiliation{Department of Physics and Astrophysics, University of Delhi, New Delhi, INDIA}
\author{B. Maheshwari}
\author{D. Choudhury}
\altaffiliation{Present Address: Tata Institute of Fundamental Research, Mumbai, INDIA}
\author{A. K. Jain}
\affiliation{Department of Physics, Indian Institute of Technology, Roorkee-247667, India}
\date{\today}
\begin{abstract}
High-spin states in the doubly-odd $^{198}$Bi nucleus have been studied by using
the $^{185,187}$Re($^{16}$O, xn) reactions at the beam energy of 112.5 MeV. $\gamma-\gamma$
coincidences, directional correlation ratio and integrated polarization asymmetry ratio were 
measured by using the INGA array with 15 Compton-suppressed clover HPGe detectors. Definite 
spins and parities were assigned to the observed levels. The high-spin structure is grouped into three 
bands (B1, B2 and B3), of which two (B1 and B2) exhibit the properties of magnetic rotation 
(MR). Tilted axis cranking calculations were carried out to explain the MR bands having large 
multi-quasiparticle (qp) configurations. The calculated results explain the bands B1 and B2 very 
nicely, confirming the shears mechanism and suggest a crossing of two MR bands in both the cases. 
The crossing is from 6-qp to 8-qp in band B1 and from 4-qp to 6-qp in band B2, a very rare finding. 
A semiclassical model has also been used to obtain the particle-hole interaction strengths below 
the band-crossing for the bands B1 and B2.

\end{abstract}
\pacs{21.10.Re; 23.20.Lv; 23.20.En; 27.80.+w }

\maketitle
\section{Introduction}
The ``Magnetic rotation (MR)", is an interesting nuclear structure phenomenon where the rotation 
of a ``magnetic top" is involved. Right conditions for such a rotation arise in weakly deformed 
configurations in which, neutron (proton) particles and proton (neutron) holes in high-j orbitals 
are coupled to each other \cite{bhu,bma}. These MR bands were observed in significant numbers in 
nuclei in the mass region $A\sim 190$ with atomic number $Z\sim 82$ \cite{amita1}, where the proton 
shell closure ensures a nearly spherical shape. The high-j proton particles in the $h_{9/2}$ and 
$i_{13/2}$ orbitals and neutron holes in the $i_{13/2}$ orbital play important roles in the Pb, Bi 
and Tl isotopes in which several good examples of MR bands were found \cite{bgb,bak,bgm,ag,hp,gz,199Bi}. 
However, proper configuration assignment to these bands remains uncertain in many of the cases due to 
incomplete knowledge of the spin and parity of the bands \cite{gz,199Bi}.

In the present work, we have investigated the high-spin states in $^{198}$Bi, where only limited information
about the high spin states was known. Apart from the ground state (T$_{1/2}$ = 10.3 min), only two other
low-lying isomeric states were known until recently. Among these, the (7$^+$) level (T$_{1/2}$ = 11.6 min)
decays by EC decay and no $\gamma$-transition from this level is known. Therefore, the excitation energy
of this state remains unknown. A 10$^{-}$ isomeric state was reported by Hagemann et al.~\cite{uh} at an 
excitation energy of 248 keV with respect to the 11.6-min isomeric state. This state was interpreted as a
member of the $\mid{\nu i^{-1}_{13/2} \otimes \pi h_{9/2};J}\rangle$ multiplet. More recently, excited states
in $^{198}$Bi were studied by Zhou et al.~\cite{xh} through $\gamma$-ray spectroscopy following
$^{187}$Re($^{16}$O,5n) reaction using six Compton suppressed HPGe detectors. The $\gamma$ rays above the
10$^-$ isomer in this nucleus were assigned from the $\gamma$-ray excitation function and a level scheme was
proposed by them from the $\gamma$-$\gamma$ coincidence measurements. In this measurement, spins and parities
were assigned to a few states only and the level scheme was known upto $\sim$ 4 MeV. The higher-lying states
in $^{198}$Bi have been interpreted by considering the coupling of the $h_{9/2}$ proton with the high spin states
in $^{197}$Pb above its 13/2$^{+}$ isomeric state. Later on, Zwartz et al.~\cite{gz} proposed three
unconnected MR-type bands in $^{198}$Bi, namely Band 1, Band 2 and Band 3, from the measurement of $B(M1)/B(E2)$
ratios. In their experiment, the $\gamma$-ray transitions were assigned in $^{198}$Bi by examining the associated BGO
calorimeter energy spectrum.  However, neither the spin-parities (J$^{\pi}$) nor the excitation energies of the
states were known and hence, the configuration of these bands also remained uncertain.

In the present work, we have carried out a detailed $\gamma$-ray spectroscopy of $^{198}$Bi upto high spins and
established the spins and parities of the excited states. This enables us to identify three band-like structures B1,
B2 and B3, of which two (B1 and B2) are proposed as MR bands. We have assigned definite configurations to both 
these bands. We further showed that each of the two bands (B1 and B2) represents a crossing of two bands with
large multi-quasiparticle (qp) configurations. In B1, the crossing is between a 6-qp and a 8-qp MR band, whereas in B2,
the crossing is between a 4-qp and a 6-qp MR band, a phenomenon very rarely observed in MR bands.

\section{Experimental Method and the Level Scheme}
The $\gamma$-ray spectroscopy of $^{198}$Bi has been studied at 15-UD Pelletron at IUAC, New Delhi, India
using the Indian National Gamma Array (INGA)~\cite{mura}. 15 Compton-suppressed clover detectors were available 
in the array at the time of the experiment. The excited states of odd-odd $^{198}$Bi were populated by 
fusion-evaporation reactions
$^{185,187}$Re($^{16}$O, xn)$^{198}$Bi at 112.5 MeV of beam energy. The target was a thick (18.5 mg/cm$^2$)
natural rhenium foil. The recoils were stopped inside the target itself. The isotopic ratio of $^{185}$Re and
$^{187}$Re in the natural rhenium is 37:63. The average beam current was $\sim 1$ pnA, limited by the count rate
of $\sim 10 000$ in a clover detector. The clover detectors were arranged in five angles with respect to
the beam direction having four clovers each at 90$^\circ$ and 148$^\circ$, two clovers each at 123$^\circ$ and
57$^\circ$, and three clovers at 32$^\circ$ angles. List mode data were recorded using $\gamma$-$\gamma$-$\gamma$
trigger. The pulse height (energy) information from individual crystals of each clover detector along with the
timing information (with respect to the master trigger) of each clover detector were recorded using conventional 
analog electronics with $CAMAC$ based data acquisition system. An in-house developed clover detector electronics 
module \cite{ingamod} was used to process energy and timing signals from Compton-suppressed clover Ge detectors.

$E_{\gamma}$-$E_{\gamma}$ matrix and $E_{\gamma}$-$E_{\gamma}$-$E_{\gamma}$ cube were constructed from the 
collected coincidence data. Inter-clover coincidences were considered to contrusct the matrix and the cube 
with energies of the $\gamma$ rays were derived using the `add-back" mode of the clovers \cite{mura,clov}. The 
random-subtracted $E_{\gamma}$-$E_{\gamma}$ matrix was primarily used for the subsequent analysis. The final 
random-subtracted matrix was constructed as follows. First a ``prompt" matrix was created with a prompt coincidence 
time-window of $\pm 40$ ns, selected from the $\gamma-\gamma$ 
time generated in the software ($\Delta$T$\gamma\gamma$) between any two of the clover detectors. Two 
additional ``random" matrices were created with time gates on the delayed parts of the $\Delta$T$\gamma\gamma$ 
spectrum. The widths of the time gates were 20-ns each and were put on both sides of the prompt peak. These two 
``random" matrices were, then, subtracted from the ``prompt" matrix to generate the final matrix. The RADWARE 
software~\cite{rw} was used for the analysis of the matrix 
and the cube. About 2.4 $\times$ 10$^8$ random- and background-subtracted true $\gamma-\gamma$ coincidence events
(obtained from the total projection of the above random-subtracted matrix after symmetrization and background
subtraction) were recorded in this experiment. The clover detectors were calibrated for $\gamma$-ray energies
and efficiencies by using $^{133}$Ba and $^{152}$Eu radioactive sources.
\begin{figure}[!]
\begin{center}
\includegraphics*[scale=0.3, angle = 0]{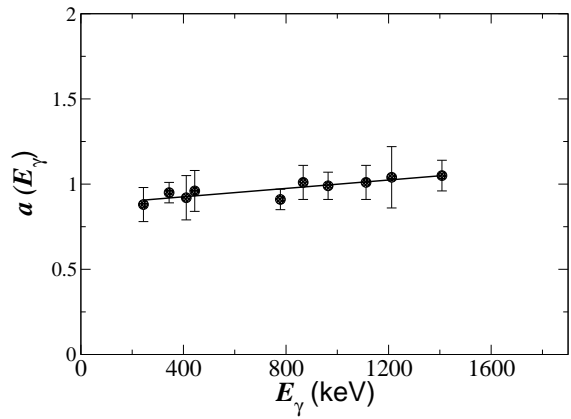}
\caption{The asymmetry correction factor $a(E_\gamma$) at different $\gamma$ energies from $^{152}$Eu source.
The solid line corresponds to a linear fit of the data.}
\label{asym}
\end{center}
\end{figure}

\begin{figure}[!]
\begin{center}
\includegraphics*[scale=0.3, angle = 0]{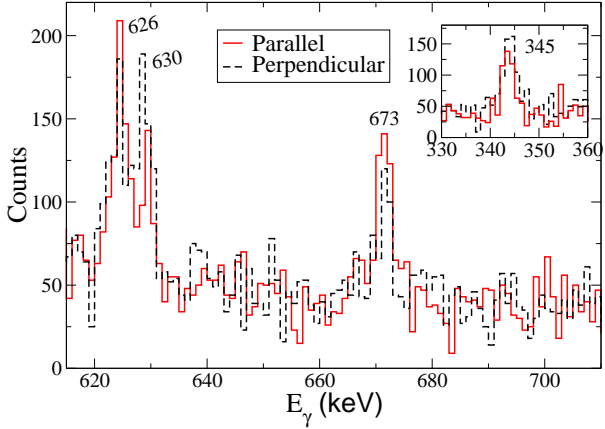}
\caption{(Color online) The perpendicular (dashed, black coloured) and parallel (solid, red coloured) components of
626-, 630- and 673-keV $\gamma$ rays in $^{198}$Bi, obtained from the IPDCO analysis in the present work. The same
for the 345-keV $\gamma$ ray is shown in the inset.}
\label{pol}
\end{center}
\end{figure}
\begin{figure*}
\begin{center}
\includegraphics*[width=17cm, angle = 0]{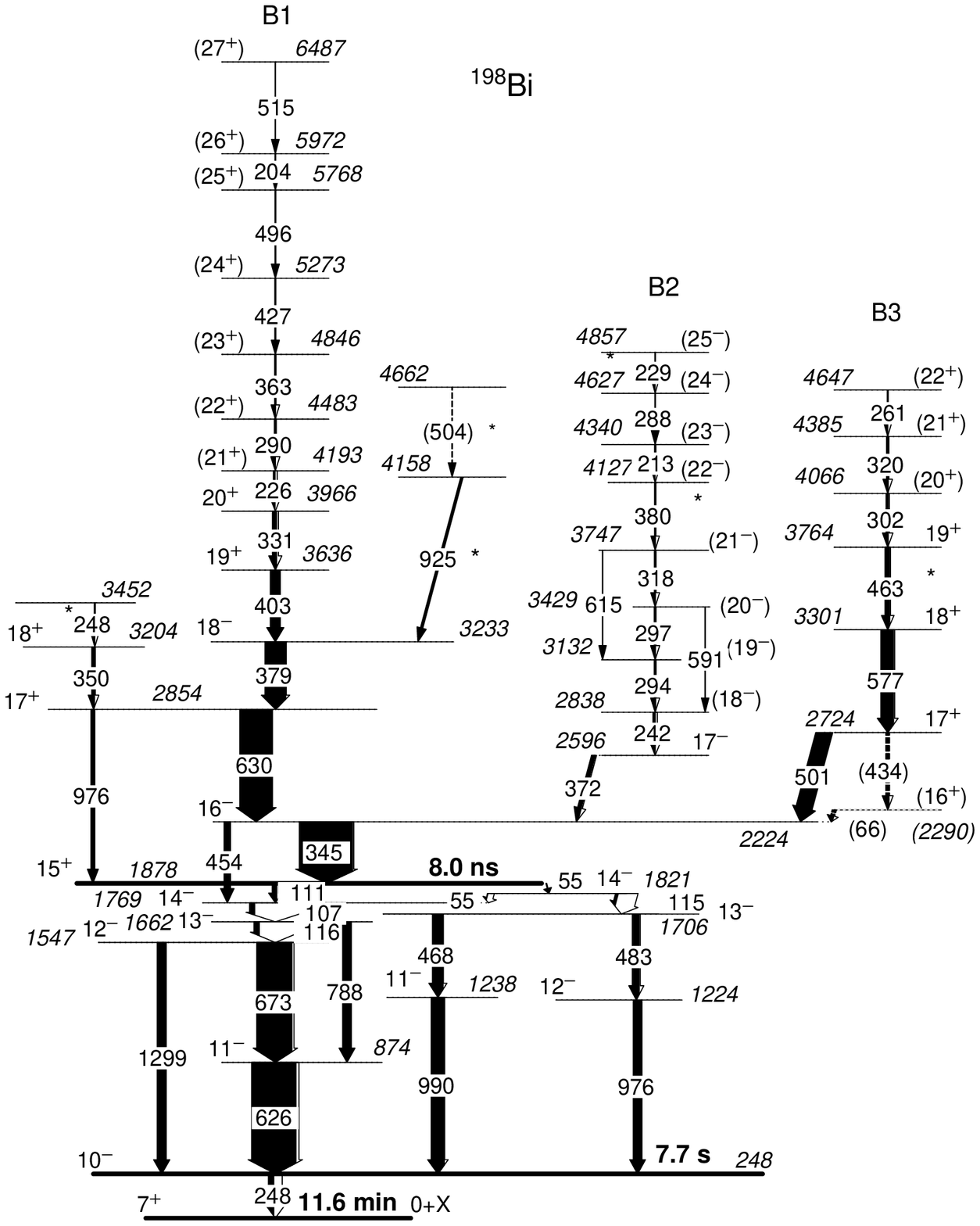}
\caption{Level scheme of $^{198}$Bi as proposed from this work. The excitation energy of the 11.6 min isomer is not
known so it has been put as 0+X. The excitation energies of all the other levels are given in reference to this isomer. 
The new $\gamma$ rays are indicated by *.}
\label{ls}
\end{center}
\end{figure*}

The multipolarities of the $\gamma$-ray transitions have been determined from the angular correlation analysis
using the method of directional correlation from the oriented states (DCO) ratio, following the prescriptions
of Kr\"{a}mer-Flecken et al.~\cite{bkm}. For the DCO ratio analysis, the coincidence events were sorted into
an asymmetry matrix with data from the 90$^\circ$ detectors ($\theta_1$) on one axis and 148$^\circ$ detectors
($\theta_2$) on the other axis. The DCO ratios (for $\gamma_1$, gated by a $\gamma$-ray $\gamma_2$ of known
multipolarity) are obtained from the intensities of the $\gamma$-rays (I$_\gamma$) at two angles $\theta_1$
and $\theta_2$, as
\begin{equation}\label{rdco}
R_{DCO} = \frac{I_{\gamma_1} ~ at~ \theta_1, ~gated ~by ~\gamma_2 ~at ~\theta_2}
               {I_{\gamma_1} ~at ~\theta_2 ~gated ~by ~\gamma_2 ~at ~\theta_1}
\end{equation}
DCO ratios for the $\gamma$ rays belonging to $^{198}$Bi were obtained by putting gates, on the transitions with 
known multipolarity, along the two axes of the above matrix. For stretched transitions, the value of R$_{DCO}$ 
would be close to unity for the same multipolarity of $\gamma_1$ and $\gamma_2$. For different multipolarities of 
$\gamma_1$ and $\gamma_2$ and for mixed multipolarity of any of these transitions, the value of R$_{DCO}$ depends 
on the detector angles ($\theta_1$ and $\theta_2$) and the mixing ratio ($\delta$).
The validity of the R$_{DCO}$ measurements was checked with the known transitions in $^{198}$Bi and with the
calculated values. In the present geometry, the calculated value of R$_{DCO}$~\cite{angcor} for a pure dipole
transition gated by a stretched quadrupole transition is 1.81 while for a quadrupole transition gated by a pure
dipole, the calculated value is 0.55. These compare well with the experimental values of 1.81(27) and 0.53(10),
respectively, for the known dipole (345.5 keV, $E1$) and known quadrupole (468.2 keV, $E2$) transitions.

Definite parities of the excited states have been assigned from the integrated polarization asymmetry
(IPDCO) ratio, as described in ref.~\cite{hp,bp1,bp2,rp}, from the parallel and perpendicular scattering
of a $\gamma$-ray photon inside the detector medium. The IPDCO ratio measurement gives a qualitative idea
about the type of the transitions ($E/M$). The IPDCO asymmetry parameter for a $\gamma$-ray transition has 
been deduced using the relation,
\begin{equation}\label{ipdco}
\Delta_{IPDCO} = \frac{a(E_\gamma) N_\perp - N_\parallel}{a(E_\gamma)N_\perp + N_\parallel},
\end{equation}
where $N_\parallel$ and $N_\perp$ are the counts for the actual Compton-scattered $\gamma$ rays in the planes
parallel and perpendicular to the reaction plane, respectively. Correction due to the asymmetry in the array 
and response of the clover segments, defined by $a(E_\gamma$) = $\frac{N_\parallel} {N_\perp}$, was checked 
using $^{152}$Eu source. The values of $a(E_\gamma$) have been obtained using the expression $a(E_\gamma$) = 
$a$ + $b(E_\gamma$). The fitting, shown in Fig.~\ref{asym}, gives the values of the constants as $a = 0.88(2)$ and
$b(E_\gamma$) = 0.00012(3). A positive value of $\Delta_{\it IPDCO}$ indicates an electric ($E$) type transition
whereas, a negative value indicates a magnetic ($M$) type transition. The low energy cut-off for the polarization
measurement in the present work was about 200 keV.

The polarization method was also verified from the $\gamma$ rays of known electric or magnetic character.
Sum gated spectra have been used for polarization measurement in the present work. The parallel ($N_\parallel$)
and perpendicular ($a(E_\gamma$)*$N_\perp$) counts for the $\gamma$ rays in $^{198}$Bi are shown in Fig.~\ref{pol}.
The 626- and 673-keV $\gamma$ rays were known to decay from 11$^-$ to 10$^-$ state and from 12$^-$ to 11$^-$
state, respectively~\cite{xh}. These transitions were known as mixed $M1+E2$ transitions. It can be clearly seen
that the counts in the parallel-scattered components are more in these two peaks than the perpendicular-scattered 
components. However, for the 345- and 630-keV $\gamma$ rays, the situation is found to be opposite. The IPDCO ratios 
for several other transitions have also been obtained in this work and together with the DCO ratio measurements,
firm spin-parity assignments could be made for several states.

An improved level scheme of $^{198}$Bi, obtained in this work, is presented in Fig.~\ref{ls}. The deduced
excitation energy, spin and parity of the excited levels and the multipolarity of the $\gamma$ rays, together
with other relevant information concerning their placement in the proposed level scheme, are summarized in
Table~\ref{tab_1}.

\begin{longtable*}{ccccccc}
\caption{\label{tab:Table1} Energies ($E_\gamma$), intensities ($I_\gamma$), DCO ratios (R$_{DCO}$),
IPDCO ratios ($\Delta_{IPDCO}$) and deduced multipolarities of the $\gamma$ rays in $^{198}$Bi.
The energies of initial states ($E_i$), spins and parities of initial ($J^\pi_i$ ) and final
($J^\pi_f$) states are also given. }\\
\hline
$E_{\gamma}$~~ &~~ $E_{i}$~~ &~~ $J^{\pi}_i$$\rightarrow$ $J^{\pi}_f$~~ &~~ $I_{\gamma}$
$^{\footnotemark[1]}$~~~ &~~~ $R_{DCO}$~~ &~~ $\Delta_{IPDCO}$~ &~ Deduced \\
(in keV) & (in keV) & & & (Err) & (Err) & Multipolarity \\
\hline
\endfirsthead
\multicolumn{7}{c}{Table~{\ref{int-198Bi}}: Continued...}\\
\hline
$E_{\gamma}$~~ &~~ $E_{i}$~~ &~~ $J^{\pi}_i$$\rightarrow$ $J^{\pi}_f$~~ &~~ $I_{\gamma}$
$^{\footnotemark[1]}$~~~ &~~~ $R_{DCO}$~~ &~~ $\Delta_{IPDCO}$~ &~ Deduced \\
(in keV) & (in keV) & & & (Err) & (Err) & Multipolarity \\
\hline
\endhead
\endfoot
\endlastfoot

(66) & (2290) & $(16^{+}) $$ \rightarrow $$     16^{-} $&  - & -  &    -    & (E1) \\

106.6(2) & 1769.0 & $14^{-} $$ \rightarrow $$     13^{-} $& 9.6(12)& 0.91(24)$^{\footnotemark[4]}$  &    -    & M1+E2 \\

110.8(4) &1878.2 & $15^{+} $$ \rightarrow $$     14^{-} $& 8.6(11)&0.98(26)$^{\footnotemark[2]}$  &  - & E1 \\

115.2(3) &1821.5 &$14^{-} $$ \rightarrow $$  13^{-} $& 5.1(7) & 1.69(46)$^{\footnotemark[3]}$ &- & M1+E2\\

115.8(2) &1662.4 &$13^{-} $$ \rightarrow $$     12^{-} $&  10.0(12)&1.05(25)$^{\footnotemark[4]}$& - & M1+E2\\

203.5(3)& 5971.3 &$(26^{+}) $$ \rightarrow $$   (25^{+}) $& 1.0(2)& - & - &  (M1)$^{\footnotemark[5]}$\\

212.9(2)& 4339.7 &$(23^{-}) $$ \rightarrow $$   (22^{-}) $&1.1(2)& - & - &  (M1)$^{\footnotemark[5]}$\\

226.2(6)&4192.5&$(21^{+}) $$ \rightarrow $$     20^{+} $&3.0(4)& - & - &  (M1)$^{\footnotemark[5]}$\\

229.2(1)&4856.7&$(25^{-}) $$ \rightarrow $$   (24^{-}) $&0.6(1)& - & - &  (M1) \\

242.2(2)&2838.2 &$(18^{-}) $$ \rightarrow $$ 17^{-} $& 4.3(5)& - & - &  (M1)$^{\footnotemark[5]}$\\

248.3(3)&3452.2&-& 0.8(1)& - & - & - \\

261.2(4)&4646.6 &$(22^{+}) $$ \rightarrow $$   (21^{+}) $&1.1(2)& - & - &  (M1)$^{\footnotemark[5]}$\\

287.8(3)&4627.5&$(24^{-}) $$ \rightarrow $$   (23^{-}) $&0.8(1)& - & - &  (M1)$^{\footnotemark[5]}$\\

290.3(2)&4482.8&$(22^{+}) $$ \rightarrow $$   (21^{+}) $&1.6(2)& - & - &  (M1)$^{\footnotemark[5]}$\\

294.1(3)&3132.3 &$(19^{-}) $$ \rightarrow $$   (18^{-}) $&2.8(3)& - & - &  (M1)$^{\footnotemark[5]}$\\

296.6(3)& 3429.3 &$   (20^{-}) $$ \rightarrow $$   (19^{-}) $&2.4(3)& - & - &  (M1)$^{\footnotemark[5]}$\\

301.8(3)&4065.5 &$   (20^{+}) $$ \rightarrow $$     19^{+} $& 4.2(6)& - & - &  (M1)$^{\footnotemark[5]}$\\

317.5(3)&3747.0 &$   (21^{-}) $$ \rightarrow $$   (20^{-}) $& 2.1(3)& - & - &  (M1)$^{\footnotemark[5]}$\\

319.8(4)&4385.3 &$   (21^{+}) $$ \rightarrow $$   (20^{+}) $& 3.2(5)& - & - &  (M1)$^{\footnotemark[5]}$\\

330.6(2)&3966.3 &$     20^{+} $$ \rightarrow $$     19^{+} $& 8.2(10)&1.01(20)$^{\footnotemark[6]}$&-0.25(10)&M1\\

345.5(1)&2223.6 &$     16^{-} $$ \rightarrow $$     15^{+} $& 100.0(30)&~~1.81(27)$^{\footnotemark[3]}$&0.18(5)&E1\\

349.9(4)&3204 &$     18^{+} $$ \rightarrow $$     17^{+} $&5.0(4)&0.94(26)$^{\footnotemark[2]}$&-&M1+E2\\

362.9(3)&4845.7 &$   (23^{+}) $$ \rightarrow $$   (22^{+}) $&1.6(2)& - & - &  (M1)$^{\footnotemark[5]}$\\

372.4(2)&2596.0 &$     17^{-} $$ \rightarrow $$     16^{-} $&7.2(6)&0.96(17)$^{\footnotemark[2]}$&-0.26(15)&M1\\

379.1(1)&3233.2 &$     18^{-} $$ \rightarrow $$     17^{+} $&39.0(20)&1.80(30)$^{\footnotemark[3]}$&0.16(7)&E1\\

379.8(1)&4126.8&$(22^{-}) $$ \rightarrow $$   (21^{-}) $&1.8(2)& - & - &  (M1) \\

402.6(4)&3635.8 &$     19^{+} $$ \rightarrow $$   18^{-} $&18.3(12)&0.91(10)$^{\footnotemark[6]}$&0.16(7)&E1\\

426.5(3)&5272.2 &$   (24^{+}) $$ \rightarrow $$   (23^{+}) $&1.3(2)& - & - &  (M1)$^{\footnotemark[5]}$\\

(434)&2724.4 &$   17^{+} $$ \rightarrow $$   (16^{+}) $&6(1)& - & - &  (M1)\\

453.7(3)&2223.6 &$     16^{-} $$ \rightarrow $$     14^{-} $&11.4(7)&0.61(12)$^{\footnotemark[4]}$&0.23(8)&E2\\

462.5(2)&3763.7 &$     19^{+} $$ \rightarrow $$     18^{+} $&8.90(7)&0.98(19)$^{\footnotemark[2]}$&-&M1\\

468.2(1)&1706.3 &$     13^{-} $$ \rightarrow $$     11^{-} $& 19.8(10)&0.53(10)$^{\footnotemark[2]}$&0.21(6)&E2\\

483.4(1)&1706.3 &$13^{-} $$ \rightarrow $$     12^{-} $&13.0(12)&1.01(25)$^{\footnotemark[2]}$&-0.26(10)&M1+E2\\

495.5(4)&5767.8 &$   (25^{+}) $$ \rightarrow $$   (24^{+}) $&1.2(2)& - & - &  (M1)$^{\footnotemark[5]}$\\

500.8(1)&2724.4 &$     17^{+} $$ \rightarrow $$     16^{-}$&30.4(15)&0.97(11)$^{\footnotemark[2]}$&0.20(11)&E1\\

504&4662 &-&1.0(1)&- & - &-\\

515.1(5)&6486.4 &$   (27^{+}) $$ \rightarrow $$   (26^{+}) $&0.8(2)& - & - &  (M1)$^{\footnotemark[5]}$\\

576.7(1)& 3301.2 &$  18^{+} $$ \rightarrow $$    17^{+} $& 23.7(15)&1.04(26)$^{\footnotemark[2]}$&-0.25(10)&M1\\

591.3(3)&3429.3 &$   (20^{-}) $$ \rightarrow $$   (18^{-}) $&0.5(1)& - & - &  (E2)$^{\footnotemark[5]}$\\

615.2(5)&3747.0 &$   (21^{-}) $$ \rightarrow $$   (19^{-}) $&0.6(1)& - & - &  (E2)$^{\footnotemark[5]}$\\

625.8(1)& 873.8 &$     11^{-} $$ \rightarrow $$     10^{-} $& 82.9(29)&1.07(16)$^{\footnotemark[2]}$&-0.14(6)&M1(+E2)\\

630.5(1)&2854.1 &$     17^{+} $$ \rightarrow $$     16^{-} $&61.1(25)&0.99(10)$^{\footnotemark[2]}$&0.21(7)&E1\\

672.8(1)&1546.6 &$     12^{-} $$ \rightarrow $$     11^{-} $&66.0(24)&0.84(12)$^{\footnotemark[2]}$&-0.24(6)&M1+E2\\

787.5(1)&1662.4 &$     13^{-} $$ \rightarrow $$     11^{-} $&16.2(11)&0.60(10)$^{\footnotemark[2]}$&0.19(8)&E2\\

924.8(2)&4158.0 &    -   &4.3(6)&-&-&-\\

975.9(2)&1223.8 &$     12^{-} $$ \rightarrow $$     10^{-} $&16.0(12)&0.54(11)$^{\footnotemark[2]}$&0.13(7)&E2\\

976.4(3)&2854.1 &$     17^{+} $$ \rightarrow $$     15^{+} $&7.1(6)&0.65(18)$^{\footnotemark[4]}$&-&E2\\

990.1(2)&1238.1 &$     11^{-} $$ \rightarrow $$     10^{-} $& 24.6(10)&1.61(29)$^{\footnotemark[2]}$&-0.12(8)&M1+E2\\

1298.7(2)&1546.6 &$     12^{-} $$ \rightarrow $$     10^{-} $&17.0(8)&0.62(14)$^{\footnotemark[2]}$&0.12(8)&E2\\

\hline
\label{int-198Bi}
\footnotetext[1]{Relative $\gamma$ ray intensities are estimated\\ from prompt spectra and
normalized to 100 for the\\ total intensity of 345.5-keV $\gamma$ rays.}
\footnotetext[2]{From 345.5 keV (E1) DCO gate;}
\footnotetext[3]{From 975.9 keV (E2) DCO gate;}
\footnotetext[4]{From 625.8 keV (M1(+E2)) DCO gate;}
\footnotetext[5]{Adopted from ref.~\cite{gz};}
\footnotetext[6]{From 379.1 keV (E1) DCO gate;}
\label{tab_1}
\end{longtable*}
\normalsize

\begin{figure*}
\begin{center}
\includegraphics*[width=12cm,keepaspectratio=true, angle = 0]{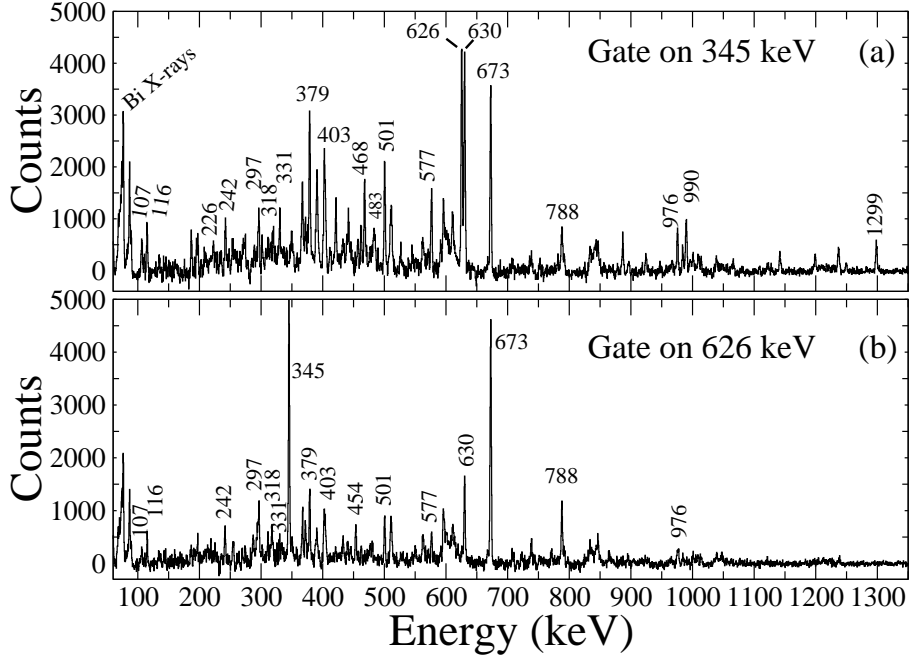}
\caption{Single-gated spectra of $^{198}$Bi gated by 345-keV (a) and 626-keV (b) transitions. The marked peaks belong
to $^{198}$Bi.}
\label{spe1}
\end{center}
\end{figure*}

Representative single-gated $\gamma$-ray spectra are shown in Figs.~\ref{spe1}(a) and Fig.~\ref{spe1}(b). These
spectra show the known $\gamma$-lines corresponding to the low-lying transitions and the $\gamma$-lines (242-,
297-, 318-, 226-, 331-, 501- and 577-keV) belonging to the proposed MR bands. Other weaker $\gamma$ lines,
those were observed in the sum-gated spectra, have been discussed later on. The 55-keV $\gamma$ ray doublet,
reported by Zhou et al.~\cite{xh}, could not be observed as the energy threshold was somewhat more than 55 keV
in the present experiment. The 223-keV $\gamma$-ray, reported in~\cite{xh}, is not observed in the present data.
So, it was not placed in the level scheme shown in Fig.~\ref{ls}.

The spins and parities of the levels up to the excitation energy of 1662 keV were assigned by Zhou et al.~\cite{xh} 
from their angular 
distribution measurements. The 107-keV transition, placed above the 1662 keV level, has been assigned as mixed 
($M1+E2$) transition in our work and, hence, the spin-parity (J$^\pi$) of 14$^{-}$ has been assigned for the 1769 
keV level.  The J$^\pi$ assignment of the 2224 keV level is crucial for the assignment of J$^\pi$ and configuration 
of the bands built on top of that level. In the previous work, J$^\pi$ of this level was tentatively assigned as
(16$^+$). This was based on the fact that 345- and 454-keV transitions were found to be dipole and quadrupole in
nature from the angular distribution measurement ~\cite{xh}. The 345-keV $\gamma$ ray was assumed to be a mixed $M1+E2$ 
transition as there was no polarization measurement. If this was the case, then the 454-keV
$\gamma$ ray turns out to be a $M2$ transition. However, no evidence for a longer halflife corresponding to such
a transition was obtained for the 2224 keV level.

In the present work, the DCO ratio measurement of the 345- and the 454-keV $\gamma$ rays indicate that they are
indeed dipole and quadrupole in nature. The polarization measurement of the 345-keV $\gamma$ ray is shown in
Fig~\ref{pol} (inset); which clearly shows that the perpendicular counts are more than the parallel counts. The
$\Delta_{IPDCO}$ value obtained for this transition suggests that it is electric type. Hence, 345-keV $\gamma$
ray is assigned as an $E1$ transition. Therefore, $J^\pi = 16^-$ is assigned for the 2224 keV level. This assignment
is well supported by the $E2$ assignment of the 454-keV $\gamma$ ray from its DCO ratio and $\Delta_{IPDCO}$
values.
Three transitions above this 2224 keV level, namely 630, 379 and 403 keV have been found to be of $E1$
type. The angular distribution measurements in the previous work \cite{xh} and the DCO ratio measurements 
in our work agree in favour of the assignment of dipole character for these transitions. The polarization 
measurements in our work show positive values of $\Delta_{IPDCO}$ for these transitions as shown in Fig.\ref{pol} 
for the 630-keV $\gamma$ ray. Therefore, the J$^\pi$ of the levels at 2854, 3233 and 3636 keV are assigned as 
17$^+$, 18$^-$ and 19$^+$, respectively. These assignments are also supported by the observed quadrupole nature of 
the 976-keV transition. The other two transitions above the 2224 keV, 16$^-$ level, namely 372- and 501-keV, have 
been found to be of $M1$ and $E1$ types from their negative and positive $\Delta_{IPDCO}$ values, respectively. 
Therefore, the band B2 is assigned to be of negative parity while the band B3 is of positive parity.

The DCO ratios of the transitions belonging to the MR bands as reported by Zwartz et al~\cite{gz} could not
be measured in the present work due to their low intensities except for a few transitions which lie towards
the bottom of the bands. These few transitions supported the assignment of ref.~\cite{gz}. So, the
multipolarity of the transitions in these bands were adopted from Zwartz et al. and are presented in
Table-I.

\begin{figure*}
\begin{center}
\includegraphics*[width=14cm,keepaspectratio=true, angle = 0]{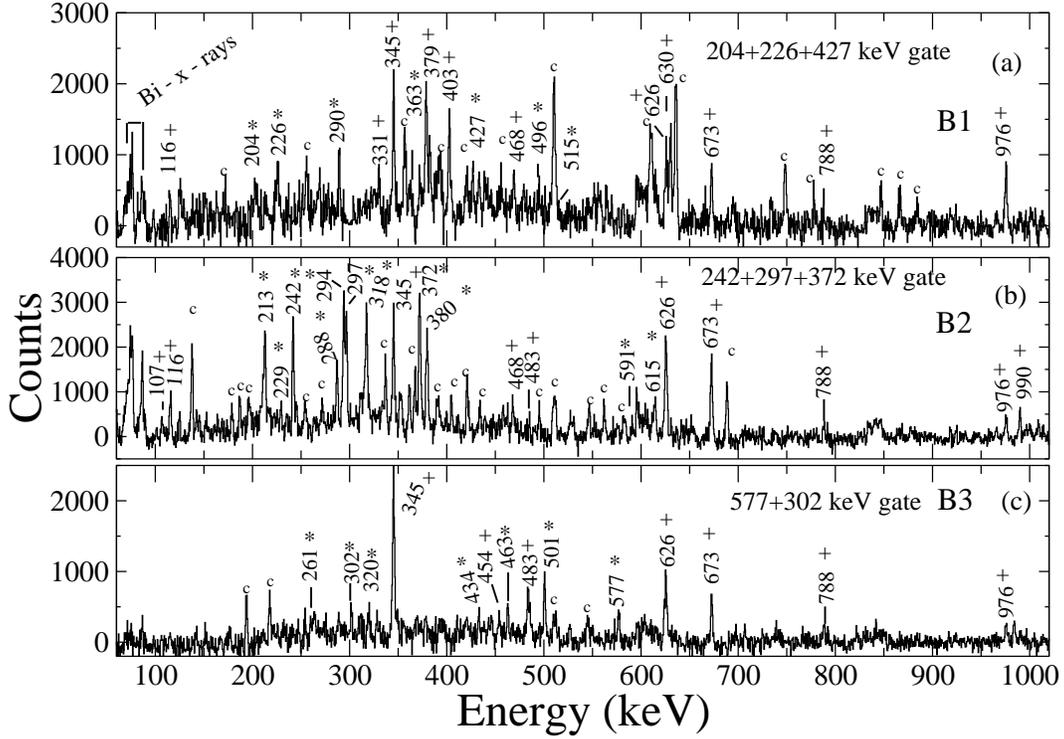}
\caption{Sum double gated spectra with gates on several in-band transitions belonging to bands B1 (a), B2 (b) 
and B3 (c) in $^{198}$Bi. The peaks marked by * and + correspond to the in-band and the lower-lying transitions,
respectively. Contaminated peaks are indicated by c.}
\label{spe2}
\end{center}
\end{figure*}

Fig.~\ref{spe2} shows three different sum coincidence spectra which were generated by summing the spectra with 
suitable gates on several in-band transitions of previously known~\cite{gz} three MR bands, B1 (Fig.~\ref{spe2}(a)), 
B2 (Fig.~\ref{spe2}(b)) and B3 (Fig.~\ref{spe2}(c)). The peaks marked by * and + in these spectra belong to
the MR bands and the previously known lower-lying transitions~\cite{xh} in $^{198}$Bi, respectively. The
sum gated spectrum of 204-, 226- and 427-keV from band B1 clearly shows all the transitions of previously
known MR band along with the known lower-lying transitions in $^{198}$Bi. In this spectrum $\gamma$ lines
belonging to the other two MR bands, B2 and B3, were not observed but the lower-lying transitions 331-, 
403-, 379-, 345-, 673-, 626-keV and other transitions of $^{198}$Bi have been observed.
Therefore, we have placed MR band B1 above the 3636-keV level.

On the other hand, the sum gated spectrum of 242-, 297- and 372-keV shows all the transitions of the band 
B2 along with the known lower-lying $\gamma$ lines in $^{198}$Bi. In this spectrum, $\gamma$ lines belonging 
to the other two bands, B1 and B3, were not observed. The 345-, 673-, 626-keV lines and other $\gamma$ lines 
of $^{198}$Bi have also been observed in this spectrum. However, the $\gamma$ lines at 331, 403, 350 and 630 
keV are absent in this spectrum. Therefore, the band B2 has been placed on top of the 2596-keV level.

Two cross-over E2 transitions have been observed in band B2: a 615-keV (from the 3747 keV level) $\gamma$
ray between the 318- and 297-keV M1 transitions, and a 591-keV (from the 3429 keV level) $\gamma$ ray between
the 297- and 294-keV M1 transitions. The $B(M1)/B(E2)$ ratios for these two levels have been found to be 6.6 (14)
and 9.2 (21) $\mu^2/(eb)^2$, respectively, which agree well with the values reported by Zwartz et al.
~\cite{gz}.

The sum-gated spectrum in Fig.~\ref{spe2}(c), obtained by putting gates on 577 keV and 302 keV, shows all the 
$\gamma$ lines of the band
B3 along with the known lower-lying $\gamma$ lines in $^{198}$Bi. In this spectrum, transitions belonging to the
other two bands, B1 and B2, have not been observed. Known $\gamma$ lines at 331, 403, 379, 350 and 630 keV were 
also not observed in this spectrum. However, 345, 673, 626 keV and other known lower-lying transitions in $^{198}$Bi
have been identified. Therefore, we have placed band B3 on top of the 2224-keV level. A strong 501-keV $\gamma$ 
line has also been observed in this spectrum which connects the band B3 with the 16$^-$ level at 2224 keV. 
There is an indication of a weak 434-keV line in the spectrum. When we gate on this 434-keV $\gamma$ peak, the 
strong lines at 577 and 463 keV are observed along with some of the low-lying known transitions but the 501-keV 
line was not observed. Also, the 434-keV line can be seen, albeit small intensity, in 577-keV gate but is not 
present in the 501-keV gate.
Therefore, we believe that the 2724-keV level decays, via the 434-keV tentative transition, to a tentatively 
assigned 16$^+$ level at 2290 keV which is the most likely band-head of the band B3.

\section{Results and Discussion}

\begin{figure}[!]
\begin{center}
\includegraphics*[width=8cm, angle = 0]{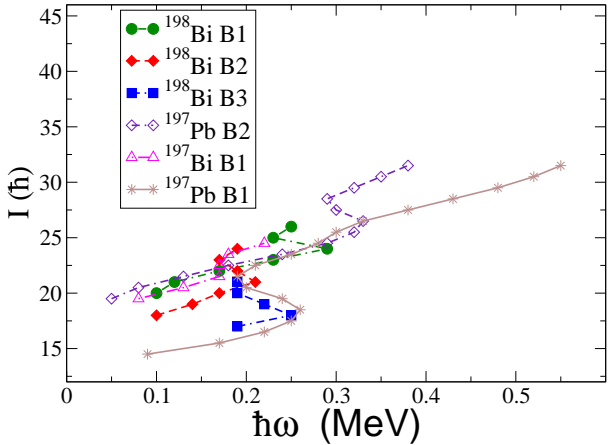}
\caption{(Color online) I (spin) vs. $\hbar\omega$ (rotational frequency) plot for the MR bands in
$^{197}$Bi~\cite{bgm},$^{198}$Bi and $^{197}$Pb ~\cite{ag}.}
\label{iw}
\end{center}
\end{figure}

The Bi isotopes with $A > 195$ are mostly near-spherical in nature. Low-lying rotational bands, 
corresponding to a well deformed shape, are observed only for the lighter Bi isotopes with neutron 
number $N \le 112$~\cite{195bi}. Therefore, the low-lying states in the odd-odd Bi nuclei with $N > 112$ 
are mostly generated by the coupling of the odd-proton and the odd-neutron in the valence orbitals near 
the Fermi energy. The $h_{9/2}$, $i_{13/2}$ and $s_{1/2}$ proton orbitals and the $i_{13/2}$, $f_{5/2}$ and 
$p_{3/2}$ neutron orbitals are available near the Fermi levels for $^{198}$Bi, which 
is weakly oblate in shape. As for example, the 2$^+$ and 7$^+$ states are generated by parallel and anti-parallel 
particle-hole couplings of $\pi h_{9/2} \otimes \nu f^{-1}_{5/2}$ orbitals. Similarly, the low-lying 10$^-$ isomer, 
observed in the odd-odd nuclei in this region, has the configuration of $\pi h_{9/2} \otimes \nu i^{-1}_{13/2}$. 
The higher-lying states may be generated by the coupling of these configurations with the two-proton or two-neutron
configurations coming from the breaking of a proton pair or a neutron pair.  For example, the important
two-neutron negative parity configurations are, 5$^-$ ($\nu i^{-1}_{13/2} p^{-1}_{3/2}$) and 9$^-$
($\nu i^{-1}_{13/2} f^{-1}_{5/2}$). Similarly, the available important proton configurations are 8$^+$
($\pi h^2_{9/2} s^{-2}_{1/2}$), 5$^-$ ($\pi h_{9/2} s_{1/2}$) and 11$^-$ ($\pi h_{9/2}i_{13/2}s^{-2}_{1/2}$),
which are observed in the even-even Pb isotopes. In fact, the $\nu i_{13/2}$ orbital, which is the highest-j orbital 
in this region, is expected to play a crucial role in most of the high-spin, multi-qp configurations.

The 15$^+$ isomer in $^{198}$Bi~\cite{xh} has been assigned a configuration of $\pi h_{9/2} \otimes
\nu i^{-2}_{13/2} p^{-1}_{3/2}$ which arises from a coupling of the 10$^-$ ($\pi h_{9/2} \otimes \nu i^{-1}_{13/2}$)
with the two-neutron 5$^-$ ($\nu i^{-1}_{13/2} p^{-1}_{3/2}$) configurations. Similarly the 16$^-$ state, the
parity of which has been assigned
in the present work, may have a configuration $\pi h_{9/2} \otimes \nu i^{-1}_{13/2} f^{-2}_{5/2}$ arising
from a coupling of the 7$^+$ ($\pi h_{9/2} \otimes \nu f^{-1}_{5/2}$) with the two-neutron 9$^-$
($\nu i^{-1}_{13/2} f^{-1}_{5/2}$) configuration. The strong transitions between the band B1 and the
16$^-$ level indicate that they have similar configurations.

As already been pointed out, the high-j  proton-particles in $\pi h_{9/2}$ and $\pi i_{13/2}$ orbitals
coupled to the neutron-holes in $i_{13/2}$ orbital ($\nu i^{-n}_{13/2}$) play important roles in the generation 
of MR bands in this region through the shears mechanism. Most
of the MR bands in odd-A Pb isotopes have been understood in terms of the 11$^-$ two-proton configuration
coupled to the $i^{-n}_{13/2}$ neutron-holes~\cite{ag}. On the other hand, the MR bands in odd-A Bi
isotopes are expected to have configurations of $\pi i_{13/2}h^{2}_{9/2} \otimes \nu i_{13/2} \nu j$ and
$\pi i_{13/2}h_{9/2}s_{1/2} \otimes \nu i_{13/2} \nu j $; where j = ($p_{3/2}, f_{5/2}$)~\cite{bgm,199Bi}.
The MR-band compilation~\cite{amita1} lists the assigned/suggested configurations for all the known MR bands.

We have plotted the angular momentum (I) vs. rotational frequency ($\hbar\omega$) for the bands B1, B2 and B3 
in Fig.~\ref{iw}. We have also compared these plots with the other MR bands in the neighboring nuclides namely, 
$^{197}$Bi and $^{197}$Pb~\cite{bgm,ag}; some similarities are noticed which are helpful in fixing the 
configurations of these bands.

To assign possible configuration for the band B1, we note the similarities of this band with the neighboring
odd-A Pb and Bi isotopes in Fig.~\ref{iw} and the fact that the band-head of this band decays strongly to the
15$^+$ isomer through a parity changing transition. As mentioned above, the configurations of the set
of levels comprising of 16$^-$, 17$^+$, 18$^-$ and 19$^+$ (the band-head of the band B1) states above the
15$^+$ isomer, may be considered as successive decoupling of neutrons (or protons) in positive and negative
parity orbitals. The fact that these states lie close in energy, excitation of protons may be ruled out.
Successive decoupling of neutrons in $i_{13/2}$ and ($p_{3/2},f_{5/2}$) orbitals, which lie close to each other
in energy, would be the most possible configuration of these states. Therefore, the configuration for the band
B1 in $^{198}$Bi may be suggested as $\pi h_{9/2} \otimes \nu i^{-2}_{13/2} (p_{3/2}, f_{5/2})^{-3}$, which
is confirmed by the TAC calculations discussed later.

A backbending has been observed in this band above 24$\hbar$ of spin corresponding to a rotational frequency
of $\hbar\omega \sim 0.29$ MeV and the band becomes a 8-qp band. This band could not be extended
much after the backbending so the details about the nature of the band after the alignment is not well known.
Most likely, the configuration of the new band after the band-crossing involves two more neutrons in the
$i_{13/2}$ orbital, which is again confirmed by the TAC calculations reported later.

Bands B2 and B3 lie at much lower excitation energies than the band B1 in $^{198}$Bi with band-head spins
and parities of 17$^-$ and (16$^+$), respectively. The band B3 (below the band-crossing) may be compared with 
the low-lying band B1 in
$^{197}$Pb, which was assigned a configuration of two-proton 11$^-$ state coupled to a neutron-hole in $i_{13/2}$
\cite{ag}. It was argued in ref.~\cite{ag} that additional two neutrons decouple at the
band-crossing frequency and this band continues with a three-neutron-hole configuration coupled to the two-proton
11$^-$ state. The I vs. $\hbar\omega$ plot for the band B1 in $^{197}$Pb is very similar to the band B3 in $^{198}$Bi.
The backbending in these two bands takes place also at the similar frequency. Considering the low energy of the
band-head, the extra odd-proton in $^{198}$Bi may be in the $h_{9/2}$ orbital. Therefore, the configuration of the
band B3 in $^{198}$Bi below the band-crossing is tentatively assigned as $\pi i_{13/2} h^{2}_{9/2} s^{-2}_{1/2} \otimes
\nu i^{-1}_{13/2}$.
Two additional neutrons decouple quickly in this band and so, after the band-crossing, it may have the
configuration of $\pi i_{13/2} h^{2}_{9/2} s^{-2}_{1/2} \otimes\nu i^{-3}_{13/2}$. However, detailed TAC
calculations are unable to reproduce the features of this band. Further, the 16$^+$ level is tentative
and the nature of the band is highly depended on the presence or absence of this state. In all likelihood, it 
appears to be a purely shell model structure based on a multi-qp configuration.

Similarly, considering the configurations reported in the neighboring odd-A nuclei and the j$^\pi$ of the band,
the band-head configuration of the band B2 in $^{198}$Bi is proposed to be $\pi h_{9/2} \otimes \nu i^{-3}_{13/2}$.
The proposed configuration is supported by the deduced g-factors, obtained from the measured B(M1)/B(E2) ratios
of the two states in this band using the geometrical model of D\"onau and Frauendorf \cite{df} and comparing them
with those calculated from the g-factors of the single-particle levels corresponding to the proposed configuration.

The Donau-Frauendorf model gives similar values of the g-factor, $g1 = 0.72(10)$ for both 20$^-$ and 21$^-$ 
states in band B2 using the experimental $B(M1)/B(E2)$ values for these two states. In these calculations, the term 
with ``g2" corresponding to the aligning quasi-particles in a band-crossing could be neglected as pointed out in 
ref. \cite{117xe}, since the $B(M1)/B(E2)$ ratios in this case are obtained only for the states below the 
band-crossing. Typical values of the intrinsic quadrupole moment $Q_o = 3.0 eb$ and the gyromagnetic factor $g_R = 0.28$ 
of the rotating core were used. The values of $g1$ obtained for the two states are very similar, implying that they 
are of same configuration. This value matches well, within uncertainties, with the value of 0.64 obtained assuming 
the assigned configuration of the band B2 before band-crossing and thereby, supporting the proposed configuration
of the band B2. 
The g-factors of the proton $h_{9/2}$ orbital $g_{h_{9/2}} = 0.786$ and the experimental value of $g_{i_{13/2}^{-3}} 
= -0.148$ for neutron $i_{13/2}^{-3}$ configuration were used \cite{gi13_2} to calculate this value. 
The TAC calculations reported in the following confirm this configuration assignment for the band B2, which also has a
band-crossing from a 4-qp to a 6-qp configuration.

\subsection{Hybrid TAC Calculations}

Tilted axis cranking (TAC) calculations have been performed to understand the band structures in $^{198}$Bi. The
hybrid TAC model as described in ref.~\cite{tac1, amita1,amita2} and references there in, has been used for the
calculations. The Hybrid version combines the best results of Woods-Saxon and Nilsson models. The single-particle
energies of the Woods-Saxon are taken and combined with the deformed part of the anisotropic harmonic oscillator.
This has the advantage of using a realistic flat-bottom potential along with the coupling between oscillator shells.
Since pairing is an important parameter in these calculations, we have calculated proton and neutron pairing gap 
parameters $\Delta_p$ and $\Delta_n$ from the ``four-point formula" using the expression given in ref.~\cite{rb}.
Assuming that the pairing gets reduced for the excited configurations, we have attenuated these parameters by 20\% 
and the proton and neutron gap parameters have been chosen to be 0.74 and 0.71 MeV, respectively.

\subsubsection{Band B1}

Calculations have been performed for several possible proton and neutron configurations of this band 
considering a 6-qp band-head with positive parity. Best results are obtained for the 6-qp configuration
$\pi h_{9/2} \otimes \nu i^{-2}_{13/2} (p_{3/2} f_{5/2})^{-3}$. After the backbending, it becomes a
8-qp configuration with an additional broken pair of neutrons in $i_{13/2}$ orbital and the neutron 
configuration changes to $\nu i^{-4}_{13/2} (p_{3/2}f_{5/2})^{-3}$. The energy minimization as a
function of deformation parameters for the 6-qp configuration gives self-consistency at the values
of $\epsilon_2=-0.083, \epsilon_4=0.003$ and $\gamma = 6.9^{\circ}$ at tilted angle $\theta = 57^{\circ}$. 
The minimized values of deformation parameters for the 8-qp configuration are $\epsilon_2 =-0.085,
\epsilon_4=0.002, \gamma=1.1^{\circ}$ at tilted angle $\theta=64^\circ$.

The spin vs. energy plot for this band is shown in Fig.~\ref{tac1}(a). It can be seen from this plot that there is
an excellent agreement of the experimental data with the calculations for the 6-qp configuration. The calculations
have been extended for the 8-qp configuration as well (shown as red dashed curve in Fig.~\ref{tac1}). It can be seen
that the spin at which the crossing of the 6-qp and the 8-qp bands takes place is well reproduced in the calculations.  
Good agreement supports the assigned band-head configuration of this band. The calculated values of $B(M1)$ are shown 
in Fig.~\ref{tac2}(a). The decreasing trend of $B(M1)$ values clearly indicates that the band B1 is an MR band. It may
also be noted that a small and constant value of $B(E2)=0.07$ eb is obtained from the calculations for this band.

In Fig.~\ref{tac3}(a), we plot the spin vs. $\hbar\omega$ values for this band and compare them with the 
calculated values. A reasonable agreement is observed before the band-crossing. The calculated results after 
the band-crossing deviates slightly from the experimental curve but the general trend is similar.

It may be noted that the other possible configurations of this band are $\pi h^2_{9/2} s_{1/2} \otimes 
\nu i^{-3}_{13/2}$ and $\pi i_{13/2} h_{9/2} s_{1/2} \otimes \nu i^{-2}_{13/2} (f_{5/2},p_{3/2})^{-1}$. The 
former configuration can be obtained from the coupling of two-proton 5$^-$ state with the 10$^-$ odd-odd 
valence-particle state along with another two decoupled neutrons in $i_{13/2}$ orbital. The later configuration 
is similar to those of band B2 in $^{197}$Pb and of the MR band in $^{199}$Bi. However, the TAC calculations for 
these bands are unable to reproduce the observed band for any of these configurations.

We may also note that the observation of a 6-qp MR band and its crossing with the 8-qp MR band is a very rare 
phenomenon, and this is only second such observation in this mass region, the first one being in $^{198}$Pb 
\cite{amita1, ag}. A similar observation has also been made in the light mass region in $^{108}$Cd 
\cite{amita1, 108cd1}.

\subsubsection{Band B2}

Since the band-head of B2 is at a lower energy than the band B1, a 4-qp configuration of
$\pi h_{9/2}\otimes \nu i^{-3}_{13/2}$ has been assumed. After the back-bending, it becomes a
6-qp configuration by breaking an additional neutron pair in $p_{3/2}$. The 6-qp configuration for
this band will then be $\pi h_{9/2} \otimes \nu i^{-3}_{13/2} p^{-2}_{3/2}$. The minimization of the deformation
parameters for 4-qp configuration gives self-consistency at the values $\epsilon_2=-0.085, \epsilon_4 =0.003$ and
$\gamma=16.4^\circ$ at an average tilted angle of $\theta=70^\circ$. After the back-bending, the minimized values
of deformation parameters for the 6-qp configuration are $\epsilon_2=-0.090, \epsilon_4=0.003, \gamma=4.5^\circ$
at an average tilted angle $\theta=65^\circ$.

\begin{figure}
\begin{center}
\includegraphics*[scale=0.3, angle = 0]{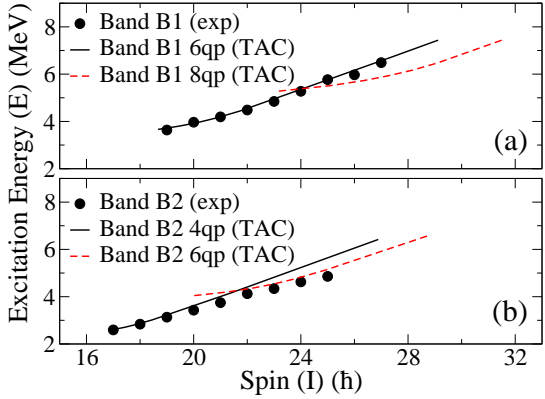}
\caption{(Color online) Spin vs. excitation energy plot for band B1 (a) and band B2 (b) in $^{198}$Bi.
The filled circles are the experimental data while the solid and the broken lines are from the TAC calculations.}
\label{tac1}
\end{center}
\end{figure}

The spin vs. energy plot for this band has been shown in Fig.~\ref{tac1}(b). Again, there is a very good 
agreement between the measured data and the calculated values. The band-crossing is also well reproduced. 
After the band-crossing at around 22$\hbar$, the experimental values follow the calculations for the 6-qp 
configuration and, there by, supporting the configurations assigned for this band both before and after the 
band-crossing.

\begin{figure}
\begin{center}
\includegraphics*[scale=0.3, angle = 0]{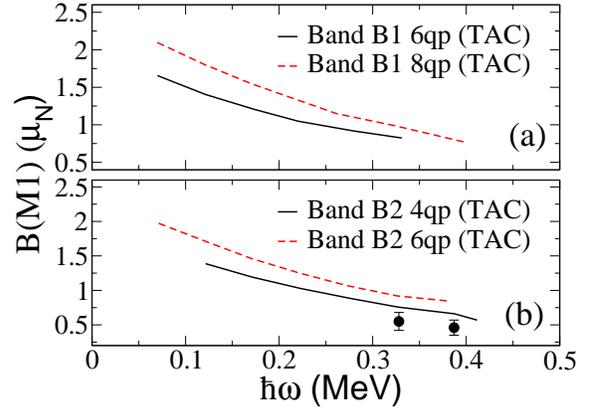}
\caption{(Color online) $B(M1)$ values, obtained from the TAC calculations, as a function of rotational frequency
$\hbar\omega$ for band B1 (a) and band B2 (b) in $^{198}$Bi. The filled circles are obtained from the measured
$B(M1)/B(E2)$ ratios and using the $B(E2)$ values from the TAC calculations for two of the states in band B2. }
\label{tac2}
\end{center}
\end{figure}

\begin{figure}
\begin{center}
\includegraphics*[scale=0.3, angle = 0]{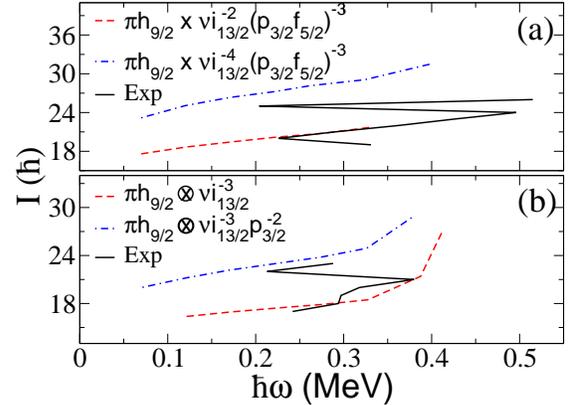}
\caption{(Color online) Spin (I) vs. rotational frequency ($\hbar\omega$) plot obtained from the TAC calculations
for band B1 (a) and band B2 (b) in $^{198}$Bi. The calculated deformation parameters for band B1 are $\epsilon_2
= -0.083, \epsilon_4=0.003, \gamma = 6.9^o$ for 6-qp configuration and $\epsilon_2 = -0.085, \epsilon_4 = 0.002,
\gamma=1.1^o$ for 8-qp configuration. The same for band B2 are $\epsilon_2 = -0.085, \epsilon_4 = 0.003, \gamma=
16.4^o$ for 4-qp configuration and $\epsilon_2 = -0.089, \epsilon_4 = 0.003, \gamma=4.5^o$ for 6-qp configuration.
The experimental values are also shown.}
\label{tac3}
\end{center}
\end{figure}

The calculated $B(M1)$ values for the 4-qp and the 6-qp configurations corresponding to the band B2 are shown 
in Fig.~\ref{tac2}(b). The decreasing values of $B(M1)$ strongly supports the MR nature of this band as well. 
As the lifetime measurements were not possible in the present experiment, so it was not possible to compare 
the calculated $B(M1)$ values with measured data. However, as mentioned earlier, two cross-over transitions 
have been observed in this band from the 20$^-$ and 21$^-$ states and the $B(M1)/B(E2)$ ratios were obtained 
for these two states from the intensities of the $\Delta I = 2$ and the $\Delta I = 1$ transitions (assuming 
mixing ratio $\delta = 0$ for the MR bands). The $B(M1)$ values can be derived from these measured ratios using 
the calculated values of $B(E2)$. The $B(M1)$ values for the two states obtained in this way are plotted in 
Fig.~\ref{tac2}(b) for comparison and an excellent agreement is observed. We have also plotted the spin (I) vs. 
$\hbar\omega$ for the band B2 in Fig.~\ref{tac3}(b) and compared with the calculated values. Reasonable agreement 
before and after the crossing supports the 4-qp and 6-qp configurations assigned for this band before and after the
band-crossing.

\subsection{Semiclassical calculations for the bands B1 and B2}

As the MR nature of the bands and their microscopic structure is well understood from the TAC calculations, the
bands B1 and B2 have been investigated in the frame work of a semiclassical model of magnetic rotation
\cite{bhu,bma,se2,se3} to get an idea about the particle-hole interaction strength. This calculation is based on a
schematic model of the coupling of two long j vectors, ({\bf $j_\pi$} and {\bf $j_\nu$}), corresponding to
the proton and neutron blades. This aims at extracting the information on the effective
interaction between the nucleons which are involved in the shears mechanism. In this model, the shears angle
($\theta$), between the two j vectors, is an important variable which can be derived using the equation
\begin{equation}
cos\theta=\frac{{I}^2-{{j}_\pi}^2-{j_\nu}^2}{2\ {j_\pi}\ {j_\nu}}
\label{sa}
\end{equation}
where, $I$ is the total angular momentum.
For the proposed configuration of $\pi h_{9/2} \otimes \nu i^{-2}_{13/2} (p_{3/2}f_{5/2})^{-3}$ for the band B1,
the $\bm{j}_\pi$ and $\bm{j}_\nu$ values are $4.5\hbar$ and $17.5\hbar$, respectively. Hence, the band-head spin
is calculated to be $18\hbar$ assuming perpendicular coupling between $\bm{j}_\pi$ and $\bm{j}_\nu$; which is
in reasonably good agreement with the observed spin of the band-head. The maximum spin for this band can be 
calculated as $22\hbar$ in this model. It can be seen from  Fig.~\ref{tac1}(a) that in the TAC calculations,
the band-crossing (between 6-qp and 8-qp bands) takes place at around the similar spin value for this band, 
indicating that the semiclassical results are consistent with the TAC calculations.
\begin{figure}
\begin{center}
\includegraphics*[scale=0.3, angle = 0]{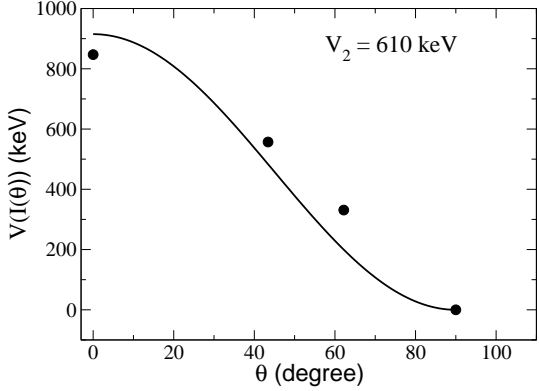}
\caption{The effective interaction between the angular momentum vectors, {\bf $j_\pi$} and {\bf $j_\nu$},
as a function of shears angle $\theta$ for the band B1 in $^{198}$Bi.}
\label{v2b1}
\end{center}
\end{figure}

According to the prescription of Macchiavelli {\it et al.}~\cite{se2}, the excitation energies of the states
in shears bands correspond to the change in the potential energy because of the re-coupling of the angular
momenta of the shears. The excitation energies of the states in the band with respect to the band-head energy
can be written as:
\begin{equation}
V(I(\theta)) = E_I - E_b = (3/2) V_2 \ cos^2{\theta}_I
\label{ve}
\end{equation}
where, $E_I$ is the energy of the level with angular momentum I, ${\theta}_I$ is the corresponding shears angle
as given in equation~\ref{sa}, $E_b$ is the band-head energy and $V_2$ is the strength of the interaction between
the blades of the shears. Therefore, $V_2$ can be calculated by using the experimentally observed energy levels
of the shears band. $V(I(\theta))$, obtained in this way for the band B1, has been plotted as a function of 
$\theta$ in Fig.~\ref{v2b1}. $V_2$ has been extracted from this plot by a fit of equation~\ref{ve}. The fitted 
curve is shown as the solid line in Fig.~\ref{v2b1}. The extracted value of $V_2$ comes out to be 610 keV, 
leading to the interaction strength per particle-hole pair ($V_2^{p-h}$) as about 122 keV considering that all 
the five possible particle-hole pairs in this configuration are contributing to the shears mechanism. This is 
somewhat less compared to a typical value of $V_2^{p-h}\sim 300$ keV observed in the Pb nuclei in this region
\cite{bma}. This possibly indicates that all the particle-hole pairs may not be contributing equally to the shears 
mechanism. A value of $V_2^{p-h}\sim 300$ keV is obtained if it is considered that the proton particles and neutron 
holes in the high-j ($\pi h_{9/2}$ and $\nu i_{13/2}$) orbitals are the only dominant contributors to the shears 
mechanism.
\begin{figure}
\begin{center}
\includegraphics*[scale=0.3, angle = 0]{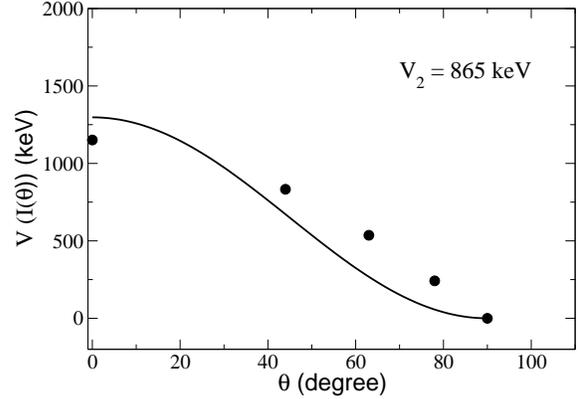}
\caption{Same as Fig.\ref{v2b1} but for band B2 in $^{198}$Bi.}
\label{v2b2}
\end{center}
\end{figure}

Similarly, for the proposed configuration of the band B2, the $\bm{j}_\pi$ and $\bm{j}_\nu$ values are $4.5\hbar$ 
and $16.5\hbar$, respectively, and therefore, the calculated minimum (band-head) and maximum spin values come
out to be $17\hbar$ and $21 \hbar$, respectively. The observed band-head spin is, again, in very good agreement 
for this band also. The maximum spin corresponding to this configuration matches well with the observed backbending 
and with the spin value of $\sim 21.5\hbar$ at which the 4-qp band crosses the 6-qp band in the TAC calculations 
(see Fig.~\ref{tac1}(b)).

After the backbending, the band has a different configuration so, in the subsequent calculations of $V(I(\theta))$
the values of $\theta$ were restricted up to the spin value of $21\hbar$ as shown in Fig.~\ref{v2b2}. The extracted
value of $V_2$ from this plot comes out to be 865 keV, this leads to the interaction strength per particle-hole
pair to be $V_2^{p-h}=288$ keV. This is in agreement with the typical values for this mass region. It may also
be noted that the configuration of this band corresponds to proton-particle and neutron-holes in high-j $\pi h_{9/2}$
and $\nu i_{13/2}$ orbitals only and hence, all the three possible particle-hole pairs contribute to the shears 
mechanism.  Therefore, it supports the conjecture that the proton-particles and neutron-holes only in high-j orbitals 
have significant roles in forming MR bands by shears mechanism in this region.

\section{\bf Summary}
We have investigated the high-spin structure of the odd-odd nucleus $^{198}$Bi by producing them in fusion-evaporation
reaction with $^{nat}$Re target and a 112.5-MeV beam of $^{16}$O using the INGA array. We have presented a new and much 
improved level scheme of this nucleus in this paper.
The DCO ratio and the polarization asymmetry measurements have also been performed to assign spins and 
parities of the levels. We have identified three band structures in this nucleus, labelled as bands B1, B2 and B3,
which lie at high excitation energies. On the basis of a comparison with similar bands in neighboring nuclei and
the TAC calculations, we conclude that the bands B1 and B2 have magnetic rotational character. One of the most
significant findings of this work is the observation of the crossing of two MR bands in B1 as well as in B2 bands.
In the band B1, the lower part of the band has a 6-qp configuration $\pi h_{9/2} \otimes \nu i^{-2}_{13/2}
(p_{3/2} f_{5/2})^{-3}$, which becomes a 8-qp configuration $\pi h_{9/2} \otimes \nu i^{-4}_{13/2} (p_{3/2}
f_{5/2})^{-3}$, after the band-crossing. Both the parts are MR in nature. Observation of such a crossing of
two MR bands having large multi-quasiparticle configuration is a very rare finding and to the best of our
knowledge, is only second such observation in this mass region. Similarly, the band B2 also undergoes a band 
crossing of two MR bands from a 4-qp configuration $\pi h_{9/2} \otimes \nu i^{-3}_{13/2}$ to a 6-qp configuration 
$\pi h_{9/2} \otimes \nu i^{-3}_{13/2} p^{-2}_{3/2}$.

Although the measurement of $B(M1)$ values are not available, the calculated $B(M1)$ values perfectly fit the 
expected behaviour of MR bands. It is, however, strongly suggested that such measurements be carried out in future. 
We have also used the semiclassical model for the configurations assigned to the bands B1 and B2 and have extracted 
the interaction strengths of the particle-hole pairs, which fits into the previously observed trend. We conclude that
the high-j proton-particle and neutron-hole pairs alone make the dominant contribution to the shears mechanism.
We also conclude that the band B3 has a pure shell model structure and is possibly not a MR band.

\section{\bf Acknowledgement}
The authors gratefully acknowledge the effort of the pelletron operators at IUAC, New Delhi for providing a good
quality of $^{16}$O beam. The help of Dr. P. Sugathan during the experiment is gratefully acknowledged. We
thank all the members of INGA collaboration for setting up the array. Fruitful discussion regarding
the polarization measurement with Prof. Maitreyee Saha Sarkar (Saha Institute of Nuclear Physics, Kolkata) is
gratefully acknowledged. We also thank Prof. A.O. Macchiavelli (Lawrence Berkeley National Laboratory, USA) for
valuable discussion. One of the authors (H. Pai) is thankful to Helmholtz International Center for FAIR for postdoctoral
fellowship and Prof. Norbert Pietralla for several scientific discussion and motivation.

\end{document}